\documentclass[aps]{revtex4}

\usepackage{graphicx}

\def\onhlf{$^1\!/\!_2$\ }

\begin{document}

\title{Symmetry and Surface Symmetry Energies in Finite Nuclei}

\author{ S.J. Lee$^{1}$ and A.Z. Mekjian$^2$}

\affiliation
{$^1$Department of Physics, Kyung Hee University, Yongin, KyungGiDo, Korea}

\affiliation
{$^2$Department of Physics and Astronomy, Rutgers University,
Piscataway, NJ 08854}

\rightline{\today}

\begin{abstract}
A study of properties of the symmetry energy of nuclei is presented based on 
density functional theory. Calculations for finite nuclei are given so that 
the study includes isospin dependent surface symmetry considerations as well 
as isospin independent surface effects. 
Calculations are done at both zero and non-zero temperature. 
It is shown that the surface symmetry energy term is the most sensitive to
the temperature while the bulk energy term is the least sensitive.
It is also shown that the temperature dependence terms are insensitive to
the force used and even more insensitive to the existence of neutron skin.
Results for a symmetry energy with both volume and surface terms are 
compared with a symmetry energy with only volume terms along the line 
of $\beta$ stability. 
Differences of several MeV are shown over a good fraction of the total
mass range in $A$.
Also given are calculations for the bulk, surface and Coulomb terms. 
\end{abstract}


\maketitle


\section{Introduction}
 
   Moderate to heavy nuclei have neutron excesses due to the growth of the Coulomb 
energy. The neutron/proton ratio $N/Z$ drifts to higher values with increasing $A$ 
with a value $\sim 1.5$ near $A = 200$. The isospin index $I = (N-Z)/A$ varies 
with neutron - proton difference and reaches a value $\sim 40/200 =0.2$ near Pb and 
somewhat higher for the very heaviest nuclei. 
For $I \ne 0$ the binding energy of nuclei then acquires an isospin 
asymmetric part with a dependence that is $I^2 A$ for the volume term. 
In finite nuclear systems a surface symmetry energy is also present and has 
a contribution with dependence $I^2 A^{2/3}$. 
The energy per particle has a well known expansion in terms of a 
liquid droplet model known as the Weizsacker mass formula 
\cite{ref1,ref2,ref3,ref4,ref5,ref6,ref7}. 
The nuclear symmetry energy along with the Coulomb energy determines the nuclear 
$\beta$ stability line. Nuclei away from the valley of $\beta$ stability will be 
explored in future rare isotope or exotic beam accelerator experiments. 
Systems with much higher neutron excess are encountered only in 
neutron stars which can have values of $I$ approaching unity. 
In neutron stars only the volume part of the symmetry energy is important. 
However, extracting the volume part of the symmetry energy in finite nuclei 
requires an analysis of both the volume symmetry and surface symmetry energy terms. 
An analysis of the division of the symmetry energy into volume and surface terms 
should lead to a better extrapolation of the symmetry energy to the limits involved 
in neutron stars. 

 The isospin dependence of binding energies arises from the underlying isospin 
structure of the nuclear force between nucleons \cite{ref7}. 
In particular the force has terms involving the isospin operator $\vec\tau$ in 
the isospin conserving form $\vec\tau_i\cdot\vec\tau_j$ which arises from 
the exchange of isovector mesons. 
Isospin symmetry is broken basically by the Coulomb force \cite{ref8,ref9}. 
A small amount of isospin breaking is also present in the nuclear interaction. 
Nuclear properties explored with heavy ion collisions result in systems that 
are not at temperature $T = 0$. 
Non-zero temperatures bring in entropy considerations in hadronic systems. 
Specifically, the Helmholtz free energy $F = E - T S$ becomes an 
important thermodynamic function with natural variables of volume $V$ and 
temperature $T$. 
Nuclear phase transitions are governed by the Helmholtz free energy 
where energy factors and entropy terms compete to determine the cluster 
distribution seen experimentally \cite{ref10,ref11}. 
Isospin properties appear in phase transitions in two component nuclear systems. 
An important example is isospin factionization \cite{ref10,ref11,ref12,ref13,ref14}
where the dilute gas phase prefers a much larger neutron excess than 
the denser liquid phase. 
The Coulomb force also plays a role in isospin factionization just as it does in 
determining the nuclear stability line \cite{ref15,ref16,ref17}. 
Similarly, the phase diagram of nuclei is a function of temperature, density 
and proton fraction which are governed by the interplay of Coulomb and 
symmetry terms \cite{ref15,ref16,ref17}. 
An extensive study of the role of isospin in heavy ion collisions can be found 
\cite{ref18,ref19}. 
Discussions of the symmetry and surface symmetry energy and how to place limits on 
the coefficients associated with them can be found in 
Ref.\cite{ref20,ref21,ref22}. 
The role of isospin asymmetry and symmetry energies in nuclear astrophysics 
are extensively discussed in Ref.\cite{ref22}. 

 This paper is devoted to a study of nuclear energies using density functional 
theory based on a Skyrme interaction. The study involves the following features.  
A). The nuclear system is finite so that both volume and surface terms appear.  
B). An $N \ne Z$ asymmetry is present so that the more general case of isospin 
$I \ne 0$ is considered with volume and surface symmetry energies both present.  
C). The proton component is charged generating a Coulomb interaction.  
D). The system is allowed to be at temperature $T \ne 0$ and entropy features are 
present. The nuclear system is therefore non-degenerate. However, the temperature is 
low enough so that an expansion around the degenerate limit can be used. 
In Sect.2 the basic relations are summarized. 
The temperature dependence of nuclear energy of finite nuclei are discussed in Sect.3
and concluded in Sect.4.
Appendix A gives some semi-analytic expressions for integrals used 
in the density functional theory.

\section{Finite temperature density functional theory and the symmetry and 
surface symmetry energies in nuclei.}
  
A density functional theory will be used to study properties of finite nuclei 
at non-zero temperature and also at temperature $T = 0$ as a limiting situation. 
We will limit the temperatures to the low temperature regime so that thermodynamic 
functions can be expanded about the degenerate limit and the $T \to 0$ can be done. 
The high temperature limit is an expansion about an ideal gas and is appropriate 
for studies of the liquid-gas phase transition \cite{ref15,ref16,ref17,prc79}.   
In this density functional approach a Skyrme Hamiltonian is used which is
\begin{eqnarray}
 H(\vec r) &=& H_B(\vec r) + H_S(\vec r) + H_C(\vec r)  \nonumber  \\
 H_B &=& \frac{\hbar^2}{2 m_p} \tau_p + \frac{\hbar^2}{2 m_n} \tau_n
                 \nonumber  \\
  & & + \frac{1}{4} \left[ t_1 \left(1 + \frac{x_1}{2}\right) 
           + t_2 \left(1 + \frac{x_2}{2}\right) \right] \rho \tau 
      - \frac{1}{4}\left[ t_1 \left(\frac{1}{2} + x_1\right)
           - t_2 \left(\frac{1}{2} + x_2\right) \right]  
         \left(\rho_p \tau_p + \rho_n \tau_n\right)
                 \nonumber  \\
  & & + \frac{t_0}{2} \left[ \left(1 + \frac{x_0}{2}\right) \rho^2 
    - \left(\frac{1}{2} + x_0\right) \left(\rho_p^2 + \rho_n^2\right) \right] 
                 \nonumber  \\
  & & + \frac{t_3}{12} \left[ \left(1 + \frac{x_3}{2}\right) \rho^2 
         - \left(\frac{1}{2} + x_3\right) \left(\rho_p^2 + \rho_n^2\right) 
        \right] \rho^{\alpha}
                 \nonumber  \\
 H_S &=& \frac{1}{16} \left[ 3 t_1 \left(1 + \frac{x_1}{2}\right)
                   - t_2 \left(1 + \frac{x_2}{2}\right) \right]
                (\vec\nabla\rho)^2
       - \frac{1}{16} \left[ 3 t_1 \left(\frac{1}{2}+x_1\right)
                   + t_2 \left(\frac{1}{2}+x_2\right) \right]
              [(\vec\nabla\rho_p)^2 + (\vec\nabla\rho_n)^2]
              \nonumber  \\
   &=& - \frac{1}{16} \left[ 3 t_1 \left(1 + \frac{x_1}{2}\right)
                   - t_2 \left(1 + \frac{x_2}{2}\right) \right]
                \rho\nabla^2\rho
       + \frac{1}{16} \left[ 3 t_1 \left(\frac{1}{2}+x_1\right)
                   + t_2 \left(\frac{1}{2}+x_2\right) \right]
              (\rho_p\nabla^2\rho_p + \rho_n\nabla^2\rho_n)
               \nonumber  \\
 H_C &=& \frac{e^2}{2} \rho_p(\vec r) 
              \int d^3 r' \frac{\rho_p(\vec r')}{|\vec r - \vec r'|}
          - \frac{3e^2}{4} \left(\frac{3}{\pi}\right)^{1/3} \rho_p^{4/3}(\vec r)
       \label{hamilt}
\end{eqnarray}
The $H(\vec r)$ has a bulk part $H_B(\vec r)$, a surface part $H_S(\vec r)$
with gradient terms and a Coulomb term $H_C(\vec r)$. 
The gradient terms are important in finite nuclei and the 
Coulomb term is important for the charged proton component.  
The $t_0$, $t_1$, $t_2$, $t_3$ and $x_0$, $x_1$, $x_2$, $x_3$ are 
parameters. 
Different choices of these parameters give rise to different Skyrme interactions. 
Here, we consider two Skyrme interactions, SKM($m^*=m$) and SLy4. 
These two Skyrme interaction have parameter sets given in Table I of Ref.\cite{prc79}.
The $m^*$ is the effective mass which is given by 
\begin{eqnarray}
 \frac{m_q}{m_q^*} &=& 1 + \frac{2 m_q}{\hbar^2} \left\{
        \frac{1}{4} \left[ t_1 \left(1 + \frac{x_1}{2}\right) 
           + t_2 \left(1 + \frac{x_2}{2}\right) \right] \rho  
      - \frac{1}{4}\left[ t_1 \left(\frac{1}{2} + x_1\right)
           - t_2 \left(\frac{1}{2} + x_2\right) \right] \rho_q \right\} 
                 \nonumber  \\
  &=& 1 + \frac{2 m_q}{\hbar^2} \left\{
       \frac{1}{16} \left[3 t_1 + (5 + 4 x_2) t_2\right] \rho
      \mp \frac{1}{8} \left[ t_1 \left(\frac{1}{2} + x_1\right)
           - t_2 \left(\frac{1}{2} + x_2\right) \right]
         \rho (2y-1) \right\}
\end{eqnarray}
At low $T$ or high density, the nearly degenerate proton and neutron Fermi gases have 
\begin{eqnarray}
 \tau_q(\vec r) &=& \frac{2m}{\hbar^2} {\cal E}_{Kq}
  = \frac{3}{5} \left(\frac{6\pi^2}{\gamma}\right)^{2/3}
   \left[\rho_q^{5/3}
   + \frac{5\pi^2 m_q^{*2}}{3\hbar^4} \left(\frac{\gamma}{6\pi^2}\right)^{4/3} \rho_q^{1/3} T^2 
   + \cdots \right]    \label{tauq}
\end{eqnarray}
The first term in square bracket is the degenerate limit and the $T^2$ term is the finite 
temperature correction. The ${\cal E}_{Kq}$  is the kinetic energy density, 
where $q=p$ for protons and $q=n$ for neutrons.  

Since our calculations are done at finite  $T$, entropy $S$ and Helmholtz free energy $F$
become important quantities with the connection to the energy $E$ through $F = E - TS$. 
The entropy density is  
\begin{eqnarray}
 T {\cal S} &=& \sum_q \frac{\hbar^2}{2 m_q^*}
   \left(\frac{6\pi^2}{\gamma}\right)^{2/3} \left[
    \frac{2\pi^2 {m_q^*}^2}{\hbar^4} \left(\frac{\gamma}{6\pi^2}\right)^{4/3} \rho_q^{1/3} T^2
   + \cdots \right]        \label{entrop}
\end{eqnarray}
for low $T$ or high density to first order in the expansion about the degenerate limit. 
The density distribution used in the evaluations presented in this work is  
\begin{eqnarray}
 \rho_q(\vec r) = \frac{\rho_{qc}}{1 + e^{(r - R)/a}}   \label{fermd}
\end{eqnarray}
In our evaluation we will first take the proton and neutron radii $R$ to be 
the same and also the diffuseness parameter $a$ to be the same.
The central density parameter $\rho_{pc}$ and $\rho_{nc}$ are determined 
to give the correct number of proton $Z$ and neutron $N$.
With this choice of the density we can integrate the energy density 
of Eq.(\ref{hamilt}) explicitly (see Appendix)
and thus the energy $E(A,Z,T)$ becomes a function of nuclear size $R$ for a
fixed value of diffuseness parameter $a$.
For each nucleus with $Z$ protons and $N$ neutrons at a temperature $T$ 
the nuclear size $R$ can be determined by minimizing the Helmholtz free 
energy.
We will also compare this situation with a case where the central
densities $\rho_{qc}$ are the same but the proton and neutron radii,
$R_p$ and $R_n$, are different.

The main quantity of interest here will be  
\begin{eqnarray}
 E(A,Z,T) 
  &=& - B(T) A + E_S(T) A^{2/3} + S_V(T) I^2 A + S_S(T) I^2 A^{2/3} 
            \nonumber \\    & &
      + E_C \frac{Z^2}{A^{1/3}} + E_{dif} \frac{Z^2}{A}
      + E_{ex} \frac{Z^{4/3}}{A^{1/3}} + c \Delta A^{-1/2}
                 \label{ldexpan}
\end{eqnarray}
where $I = (N-Z)/A = (A - 2Z)/A$.
The $E_{dif}$ and $E_{ex}$ are the coefficients for the diffuseness correction 
and the exchange correction to the Coulomb energy.  
For the pairing correction with constant $\Delta$, $c = +1$ for odd-odd
nuclei, 0 for odd-even nuclei, and $-1$ for even-even nuclei.
The above formula at $T =0$ is the well known Weizacker semiempirical mass 
formula \cite{ref1,dan9,dan11} studied extensively by Myers and 
Swiatecki \cite{dan10,dan14,dan15,dan16}. Early studies excluded the surface 
symmetry term $S_S$ and only the surface term $E_S$ was included. The values of the 
coefficients as found in textbooks such as Ref.\cite{ref7} 
are $B(0) \approx 16$, $E_S(0) \approx 17$, and $S_V(0) \approx 24$ in MeV. 
The ratio $E_S/B$ of surface to bulk energy at $T=0$ is very close to unity. 

With regard to the volume symmetry term about \onhlf
the numerical value of the coefficient $S_V(T=0)$ comes from kinetic energy 
considerations for two degenerate Fermi gases of proton and neutrons. 
Specifically, the kinetic energy contribution is simply related to the Fermi 
energy $E_F$ as 
\begin{eqnarray}
 S_V(0)_{kin} = \frac{1}{3} E_F \approx 12 {\rm MeV}
\end{eqnarray}
The other \onhlf of the symmetry energy coefficient arises from interaction 
terms.  
For systems with a neutron excess, the neutrons and protons experience 
interaction potentials. In an independent particle model the average 
potential, (called a Lane potential, see Ref.{\cite{ref7}), 
felt by a neutron or proton differ. The potential is written in the form
\begin{eqnarray}
 V = V_0 + \frac{1}{2} t_z \frac{N-Z}{A} V_1
\end{eqnarray}
with $t_z = 1/2$ for neutrons and $t_z = -1/2$ for protons. 
A total potential energy can be obtained by summing the one body potential 
energy and taking \onhlf of the result to give
\begin{eqnarray}
 V_{pot} = \frac{1}{2} (N+Z) V_0 + \frac{1}{8} \frac{N-Z}{A} V_1 N
    - \frac{1}{8} \frac{N-Z}{A} V_1 Z
  = \frac{1}{2} A V_0 + \frac{1}{8} \frac{(N-Z)^2}{A} V_1
\end{eqnarray}
A depth $V_1 = 96$ MeV gives $S_V(0)_{pot} = V_1/8 = 12$ MeV, 
the remaining part of the symmetry energy coefficient $S_V(0) = 24$ MeV.
More recently, the importance of a surface symmetry 
term has been noted. Several calculations at $T=0$ have been presented 
regarding this term \cite{ref20,ref21,ref22}. 
Considerable variation in the ratio of the $T$ independent surface to volume 
symmetry energy with the difference $R_n - R_p$ were noted \cite{ref22}.
Here our focus will be on the $T$-dependent features which arises 
from the kinetic energy terms and effective mass terms.

The Coulomb energy $E_{Coul}$ for a proton distribution of Eq.(\ref{fermd}) 
(see p.145 and p.160 of Ref.\cite{ref7}) is
\begin{eqnarray}
 E_{Coul} &=& \int d^3 r H_C(\vec r) 
   = \frac{3}{5} e^2 \frac{Z^2}{R} \left[1
       - \left(\frac{7\pi^2}{6}\right)\left(\frac{a}{R}\right)^2\right]
     - \frac{3}{4} e^2 \left(\frac{3}{\pi}\right)^{1/3}
         \rho_p^{4/3} \frac{4\pi}{3} R^3    \nonumber  \\
  &=& \frac{3}{5} \frac{e^2}{r_0} \frac{Z^2}{A^{1/3}}
     - \frac{7}{10} e^2 \pi^2 \frac{a^2}{r_0^3} \frac{Z^2}{A}
     - \frac{3}{4} \frac{e^2}{r_0} \left(\frac{3}{2\pi}\right)^{2/3}
            \frac{Z^{4/3}}{A^{1/3}}
\end{eqnarray}
with $R = r_0 A^{1/3}$. 
For $a = 0.53$ fm and $r_0 = 1.25$ fm, $E_C = 0.6912$, $E_{dif} = -1.43081$, 
and $E_{ex} = - 0.5278064$ in MeV unit.

\section{Temperature dependence of Energy of Finite Nuclei}

To obtain the temperature dependence of the various coefficients, the Helmholtz 
free energy is minimized 
for various nuclei along $\beta$ stability line with $Z$ protons 
and $N$ neutrons from $^{20}$Ne to $^{208}$Pb.
The Eq.(\ref{ldexpan}) is then used to obtain the expansion coefficients 
of energy.
The results for two different Skyrme interactions,
SKM($m^*=m$) with no effective mass and SLy4 with a density dependent 
effective mass of $m^*/m = 0.7$ at nuclear matter density \cite{prc79,skmst}, 
are presented in Table \ref{encoef}.

\begin{table}
\caption{Energy coefficient minimizing free energy in MeV unit.
The rows labeled with ``$T^2$ term'' are the expansion of the coefficient
of the explicit $T^2$-dependent term of the kinetic energy coming from 
the $T^2$ term in Eq.(\ref{tauq}), the kinetic energy term, 
and the rows labeled with ``$T$-indep.'' are the expansion of the remainders 
in $E(A,Z,T)$.
The rows labeled with ``$T^2(R_n\ne R_p)$'' are for $R_n\ne R_p$ and
are explained in text.
All other rows are for $R_n=R_p$.
}
    \label{encoef}
\begin{tabular}{l|rrrr|rrrrr}
\hline
             &  SKM&($m^*=m$) \ \ \ \  &            &        
             &  SLy4      &        &            &           \\
\hline
 $T$ (MeV)   &  0 \ \     &  1 \ \     &  2 \ \     &  3 \ \   
             &  0 \ \     &  1 \ \     &  2 \ \     &  3 \ \     \\
\hline
 $B(T)$      &  15.310  &  15.270  &  15.152  &  14.955  
             &  15.308  &  15.296  &  15.264  &  15.217  \\
 \ \ $T$-indep.& 15.310 &  15.310  &  15.310  &  15.310  
             &  15.308  &  15.308  &  15.308  &  15.313  \\
 \ \ $T^2$ term &       &\ 0.039416&\ 0.039392&\ 0.039448  
             &          &\ 0.011491&\ 0.011169&\ 0.010581   \\
 \ \ $T^2(R_n\ne R_p)$ & & 0.039465 & 0.039599 & 0.039436 
             &          & 0.011364 & 0.011257 &  0.010709  \\
\hline
 $E_S(T)$    &  18.303  &  18.804  &  20.324  &  22.905  
             &  20.008  &  20.558  &  22.249  &  25.205  \\
 \ \ $T$-indep.& 18.303 &  18.304  &  18.311  &  18.340  
             &  20.008  &  20.008  &  20.024  &  20.099  \\
 \ \ $T^2$ term &       &  0.50052 &  0.50314 &  0.50729 
             &          &  0.54997 &  0.55604 &  0.56734 \\
 \ \ $T^2(R_n\ne R_p)$ & & 0.50037 &  0.50240 &  0.50738 
             &          &  0.55043 &  0.55572 &  0.56694 \\
\hline
 $S_V(T)$    &  19.685  &  20.108  &  21.415  &  23.673  
             &  31.113  &  31.663  &  33.403  &  36.656  \\
 \ \ $T$-indep.& 19.685 &  19.685  &  19.699  &  19.745  
             &  31.113  &  31.114  &  31.152  &  31.339  \\
 \ \ $T^2$ term &       &  0.42287 &  0.42898 &  0.43643 
             &          &  0.54823 &  0.56287 &  0.59079 \\
 \ \ $T^2(R_n\ne R_p)$ & & 0.43200 &  0.43658 &  0.44579 
             &          &  0.55591 &  0.56828 &  0.59845 \\
\hline
 $S_S(T)$    & --33.176 & --35.222 & --41.563 & --52.538 
             & --41.035 & --43.639 & --51.805 & --66.974 \\
 \ \ $T$-indep.&--33.176& --33.167 & --33.220 & --33.430 
             & --41.035 & --41.045 & --41.191 & --41.992 \\
 \ \ $T^2$ term &       & --2.0551 & --2.0858 & --2.1230 
             &          & --2.5948 & --2.6534 & --2.7758 \\
 \ \ $T^2(R_n\ne R_p)$ & &--2.0790 & --2.1088 & --2.1460 
             &          & --2.5958 & --2.6531 & --2.7905 \\
\hline
\end{tabular}
\end{table}
The central density $\rho_c$ for SKM($m^*=m$) has a range 
$0.14595 \sim 0.16697$ fm$^{-3}$ at $T = 0$,
$0.14556 \sim 0.16618$ fm$^{-3}$ at $T = 1$ MeV,
$0.14446 \sim 0.16369$ fm$^{-3}$ at $T = 2$ MeV,  
$0.14254 \sim 0.15944$ fm$^{-3}$ at $T = 3$ MeV,
$0.13973 \sim 0.15290$ fm$^{-3}$ at $T = 4$ MeV, and 
$0.13586 \sim 0.14327$ fm$^{-3}$ at $T = 5$ MeV 
with maximum for $^{20}_{10}$Ne$_{10}$ and minimum for $^{208}_{\ 82}$Pb$_{126}$.
The central density $\rho_c$ for SLy4 has a range 
$0.14286 \sim 0.14886$ fm$^{-3}$ at $T = 0$,
$0.14223 \sim 0.14799$ fm$^{-3}$ at $T = 1$ MeV,
$0.14017 \sim 0.14542$ fm$^{-3}$ at $T = 2$ MeV,
$0.13664 \sim 0.14091$ fm$^{-3}$ at $T = 3$ MeV,
$0.13117 \sim 0.13392$ fm$^{-3}$ at $T = 4$ MeV 
with minimum for $^{208}_{\ 82}$Pb$_{126}$ and maximum for $^{64}_{30}$Zn$_{34}$. 
At $T = 5$ MeV, it has a range $0.12302 \sim 0.12519$ with maximum
for $^{116}_{\ 50}$Sn$_{66}$ and minimum for $^{208}_{\ 82}$Pb$_{126}$ 
for $Z = 30 \sim 82$ region and $\rho_c = 0.09748$ for $^{20}_{10}$Ne$_{10}$ 
and $\rho_c = 0.11736$ for $^{40}_{20}$Ca$_{20}$.
As the temperature $T$ increases, the central density $\rho_c$ which 
minimizes the free energy $F$ decreases
and thus the nuclear size $R$ becomes larger.
This indicates that no nucleus can be bound with a minimum value of $F$
at high temperature due to the entropy which increases as $R$ increases. 
For $^{20}_{10}$Ne$_{10}$ with SLy4 parameter,
there was no $\rho_c$ or $R$ value which minimizes $F$ 
at $T \buildrel>\over\sim 5.3$ MeV.
The density $\rho_c$ which minimizes the free energy $F$ is sensitive to 
the interaction used.

The results of Table \ref{encoef}, from the values of ``$T$-indep.'' 
and ``$T^2$ term'' which are approximately $T$ independent, 
can be summarized approximately at low $T$ as 
\begin{eqnarray}
 B(T) &=& 15.310 - 0.0394 T^2 
                    \nonumber \\
 E_S(T) &=& 18.303 + 0.501 T^2 
                    \nonumber \\
 S_V(T) &=& 19.685 + 0.423 T^2 
                    \nonumber \\
 S_S(T) &=& -33.176 - 2.055 T^2 
                    \label{expskm}
\end{eqnarray}
in MeV unit for SKM($m^*=m$) and
\begin{eqnarray}
 B(T) &=& 15.308 - 0.0115 T^2 
                    \nonumber \\
 E_S(T) &=& 20.008 + 0.550 T^2 
                    \nonumber \\
 S_V(T) &=& 31.113 + 0.548 T^2 
                    \nonumber \\
 S_S(T) &=& -41.035 - 2.595 T^2 
                    \label{expsly4}
\end{eqnarray}
in MeV unit for SLy4.
When the Helmholtz free energy $F$ is minimized, 
the $T$ dependence of energy $E(A,Z,T)$ at low $T$ becomes
\begin{eqnarray}
 E(A,Z,T) &=& - (15.31 - 0.04 T^2) A + (18.30 + 0.50 T^2) A^{2/3} + (19.69 + 0.42 T^2) I^2 A
            \nonumber \\    & &
       - (33.18 + 2.06 T^2) I^2 A^{2/3}   
      + E_C \frac{Z^2}{A^{1/3}} + E_{dif} \frac{Z^2}{A}
      + E_{ex} \frac{Z^{4/3}}{A^{1/3}} + c \Delta A^{-1/2}   \label{etskm}  \\
 E(A,Z,T) &=& - (15.31 - 0.11 T^2) A + (20.01 + 0.55 T^2) A^{2/3} + (31.11 + 0.55 T^2) I^2 A
            \nonumber \\    & &
       - (41.04 + 2.59 T^2) I^2 A^{2/3} 
      + E_C \frac{Z^2}{A^{1/3}} + E_{dif} \frac{Z^2}{A}
      + E_{ex} \frac{Z^{4/3}}{A^{1/3}} + c \Delta A^{-1/2}   \label{etsly4}
\end{eqnarray}
for SKM($m^*=m$) and SLy4 respectively in MeV unit.

The temperature dependence of energy Eq.(\ref{ldexpan}) 
and Eqs.(\ref{expskm}) - (\ref{etsly4}) comes
from the $T$ dependent second term in Eq.(\ref{tauq}), the kinetic energy. 
Comparing Eqs.(\ref{tauq}) and (\ref{entrop}), we can see the entropy $T S$ 
term is twice that of the $T$ dependent part of the kinetic energy with 
the effective mass ${\cal E}^*_{Kq} = \frac{\hbar^2}{2 m^*_q} \tau_q$.
This means that the $T$ dependence of energy originate mainly from
the $T$ dependence of the entropy and  
the $T$ dependence of the free energy $F = E - TS$ is the same as the $T$ 
dependence in $E$ with an opposite sign at low $T$.
The entropy can be approximated as
\begin{eqnarray}
 S &=& T \left(0.079 A + 1.00 A^{2/3} + 0.85 I^2 A - 4.11 I^2 A^{2/3}\right)
        \label{efskm}
\end{eqnarray}
for SKM($m^*=m$) and
\begin{eqnarray}
 S &=& T \left(0.22 A + 1.10 A^{2/3} + 1.10 I^2 A - 5.19 I^2 A^{2/3}\right)
        \label{efsly4}
\end{eqnarray}
for SLy4 in MeV unit. 
The volume term of Eqs.(\ref{etskm}) - (\ref{efsly4})
has the smallest coefficient and the surface symmetry term has 
the largest coefficient among four terms.
The minimum free energy $F = E - T S$ of finite nuclei at low $T$ becomes
\begin{eqnarray}
 F(A,Z,T) &=& - (15.31 + 0.04 T^2) A + (18.30 - 0.50 T^2) A^{2/3} + (19.69 - 0.42 T^2) I^2 A
            \nonumber \\    & &
       - (33.18 - 2.06 T^2) I^2 A^{2/3}   
       + E_C \frac{Z^2}{A^{1/3}} + E_{dif} \frac{Z^2}{A}
       + E_{ex} \frac{Z^{4/3}}{A^{1/3}} + c \Delta A^{-1/2}   \label{ftskm}  \\
 F(A,Z,T) &=& - (15.31 + 0.11 T^2) A + (20.01 - 0.55 T^2) A^{2/3} + (31.11 - 0.55 T^2) I^2 A
            \nonumber \\    & &
       - (41.04 - 2.59 T^2) I^2 A^{2/3} 
       + E_C \frac{Z^2}{A^{1/3}} + E_{dif} \frac{Z^2}{A}
       + E_{ex} \frac{Z^{4/3}}{A^{1/3}} + c \Delta A^{-1/2}   \label{ftsly4}
\end{eqnarray}
for SKM($m^*=m$) and SLy4 respectively in MeV unit.

The results of Table \ref{encoef} and Eqs.(\ref{expskm}) - (\ref{etsly4}) show 
the $T$ dependence of the bulk, surface, symmetry and surface symmetry energy 
from $T=0$ MeV to $T=3$ MeV. 
At $T=0$ the $F$ and $E$ are the same and thus the minimum of each of
them are also the same. 
Eq.(\ref{tauq}) shows a characteristic $T^2$ 
dependence coming from a nearly degenerate Fermi gases of protons and neutrons. 
The bulk and surface terms are similar for the two Skyrme interactions. 
However, a comparison of the symmetry term and surface symmetry term at $T=0$
are quite different showing a sensitivity to the force used. 
The large variation of symmetry energy depends on the force used, 
which is also shown in Ref.\cite{npa818} at zero temperature.
(In Ref.\cite{npa818} $a_a^S=E_V^2/E_S$ is given instead of $E_S$.)
The SLy4 interaction has a much larger coefficients than the SKM($m^*=m$) interaction. 
Most of the difference comes from the $T$ independent part. The $T^2$ dependent
term of symmetry energy is much less sensitive to the force used.
The volume symmetry and surface symmetry terms at $T=0$ are also sensitive
to the difference in neutron - proton radii \cite{ref22}. However,
in Ref.\cite{prl85} an investigation of the neutron - proton radii difference
in $^{208}$Pb using a Skyrme-Hartree-Fock model shows a small difference
of $R_n - R_p = 0.16$ fm compared to nuclear size of about 7 fm.

The results show that 
the magnitude of $B$ becomes smaller as $T$ increases while
the other coefficients become larger.
The bulk energy $B$ is insensitive to the temperature
having a small coefficient for the $T^2$ term.
The surface energy $E_S$ and the volume symmetry energy $S_V$ have larger
temperature dependences which are of the same order.
The surface symmetry energy $S_S$ is the most sensitive to the temperature
having a large coefficient in front of $T^2$.
The $T$ dependent terms are insensitive to the force used.
Most of the dependence on the force used appears in the $T$ independent part 
of energy (see Eq.(\ref{hamilt})).
This fact shows that the $T$ dependence comes from the kinetic
energy including an effective mass term, and only the effective mass part has 
a dependence on force used. The interaction part which is sensitive to the
force used is insensitive to the temperature.
The central density $\rho_c$ or the nuclear size $R$ which minimizes
the free energy $F$ is sensitive to the interaction used.
From Table \ref{encoef}, we can see there is a small extra $T$ dependence 
beside the dominant $T^2$ dependence given by Eqs.(\ref{expskm})
and (\ref{expsly4}). 
Including this small extra T dependence, Eq.(\ref{expskm}) 
and Eq.(\ref{expsly4}) have small corrections that are as follows.
By extracting $T$ dependence up to $T^3$ directly
from $B(T)$, $E_S(T)$, $S_V(T)$, and $S_S(T)$  
in Table \ref{encoef} itself,
for SLy4, the $T$ dependences of Eq.(\ref{expsly4}) are now 
\begin{eqnarray}
 B(T) &=& 15.30754 + 0.00122333 T - 0.0138600 T^2 + 0.00114667 T^3  \nonumber \\
 E_S(T) &=& 20.00770 + 0.0226783 T + 0.507110 T^2 + 0.02095167 T^3  \nonumber \\
 S_V(T) &=& 31.11260 + 0.0619383 T + 0.434450 T^2 + 0.05360167 T^3  \nonumber \\
 S_S(T) &=& -41.03514 - 0.304690 T - 2.059045 T^2 - 0.2404950 T^3  \label{extr}
\end{eqnarray}
The coefficients for SKM($m^*=m$) have the similar $T$ dependence but
less sensitive than for SLy4 case. 
Specifically, Eq.(\ref{expskm}) is now modified to
\begin{eqnarray}
 B(T) &=& 15.30963 - 0.0004933 T - 0.0386350 T^2 - 0.000201667 T^3  \nonumber \\
 E_S(T) &=& 18.30316 + 0.0063483 T + 0.487620 T^2 + 0.00720167 T^3  \nonumber \\
 S_V(T) &=& 19.68525 + 0.0024133 T + 0.408910 T^2 + 0.01111667 T^3  \nonumber \\
 S_S(T) &=& -33.17567 - 0.011930 T - 1.977970 T^2 - 0.0564600 T^3  \label{extrskm} 
\end{eqnarray}
This extra $T$ dependence is due to the different density (central density 
$\rho_c$) which minimizes $F$ for different $T$ and causes the small variation 
of the coefficients in Table \ref{encoef} as $T$ changes  
and the differences between Eqs.(\ref{expsly4}) and (\ref{extr}) 
and between Eqs.(\ref{expskm}) and (\ref{extrskm}). 
The density $\rho_c$ is sensitive to the interaction used 
as mentioned before.

Here we used the same size of proton and neutron distribution, $R_p = R_n$.
In this case the system has no neutron skin. 
The existence of neutron skin may affects the surface dependence of energy.
However since we are interested in the temperature dependence of energy
we examined the $T^2$ dependent term in kinetic energy (Eq,(\ref{tauq})) 
by evaluating with different values of $R_p$ and $R_n$ which are determined 
by requiring the correct number of $Z$ and $N$ with the same central 
density of $\rho_{pc} = \rho_{nc} = \rho_c/2$ where $\rho_c$ is 
the total central density of the results minimizing $F$ with $R_p=R_n$.
The results are shown in Table \ref{encoef} labeled with ``$T^2 (R_p\ne R_n)$''
which are very close to the values of ``$T^2$ term''.
This shows that the effect of the different size of neutron and proton
distribution is much smaller than the effect of different force parameter set.
For the case of $R_p=R_n$ the most asymmetry occurs in the central region
while it occurs at the surface region for the case of $\rho_{pc} = \rho_{nc}$.
If the overall asymmetry effects of these two extreme cases are the same then 
the total symmetry energy is insensitive to the existence of a neutron skin.
The Weizacker expansion of Eq.(\ref{ldexpan}) cannot distinguish between
effects coming from different central densities and different radii.
This result is shown by comparing the values in the rows of ``$T^2$ term''
and ``$T^2 (R_p\ne R_n)$'' for each case in Table \ref{encoef}.
Of course the $T$ independent part may have a larger effect for different values
of $R_p$ and $R_n$ than the $T^2$ dependent part.
A discussion of the surface to volume symmetry energy at $T = 0$ 
can be found in Ref.\cite{ref22}.
The energy expansion coefficients for some other Skyrme parameter sets
with various effective masses are also shown in Table \ref{encoef2}. 
In Tables \ref{encoef} and \ref{encoef2} we compared the results 
for the parameter sets with wide range of effective masses
from $m^*/m = 1.0$ to 0.577. Eqs.(\ref{hamilt}) - (\ref{tauq}) show
the temperature dependence of energy comes mostly from the effective mass.
Tables \ref{encoef} and \ref{encoef2} show that the above discussions
about the qualitative behavior of the temperature dependence of energy 
coefficients are independent to the force parameter sets used.  
Of course the quantitative values of the $T$ dependences depend on
the Skyrme parameters used.
These tables show that the effect of the effective mass on the $T$ dependence 
of the energy is most visible in the $T^2$ term of the volume energy $B(T)$. 
For smaller effective mass the $T^2$ dependent term of $B(T)$ has the tendency
of becoming smaller which can be understood by Eq.(\ref{tauq}).
The other energy coefficients do not show any specific pattern of the
dependence on the effective mass.

\begin{table}
\caption{Same as Table \ref{encoef} 
but for different Skyrme interaction sets of SkT8 \cite{skt8}, 
SkM$^*$ \cite{skmst}, and SkI3 \cite{ski3}
with the effective mass $m^*/m$ of 0.833, 0.79, and 0.577 respectively.
}
    \label{encoef2}
\begin{tabular}{l|rrr|rrr|rrrrrr}
\hline
             &  SkT8      &     &      
             &  SkM$^*$     &   &     
             &  SkI3     &      &      \\ 
\hline
 $T$ (MeV)   &  0 \ \     &  1 \ \     &  2 \ \  
             &  0 \ \     &  1 \ \     &  2 \ \  
             &  0 \ \     &  1 \ \     &  2 \ \  \\ 
\hline
 $B(T)$      &  14.940 &  14.922  &  14.869   
             &  15.127 &  15.108  &  15.051   
             &  15.102 &  15.102  &  15.104  \\ 
 \ \ $T$-indep.&14.940 &  14.940  &  14.942    
             &  15.127 &  15.127  &  15.127       
             &  15.102 &  15.102  &  15.104  \\ 
 \ \ $T^2$ term &       &\ 0.01853 &\ 0.01810    
             &          &\ 0.01921 &\ 0.01911     
             &          &\ 0.00048 &\ --0.00002  \\ 
 \ \ $T^2(R_n\ne R_p)$ & & 0.01851 &  0.01809   
             &          & 0.01918 & 0.01909      
             &          & 0.00044 & --0.00005  \\ 
\hline
 $E_S(T)$    &  21.684 &  22.266  &  24.056      
             &  18.756 &  19.296  &  20.947       
             &  22.386 &  22.963  &  24.735   \\ `
 \ \ $T$-indep.& 21.684 &  21.685  &  21.704     
             &  18.756  &  18.757  &  18.770      
             &  22.386  &  22.387  &  22.404   \\ 
 \ \ $T^2$ term &       &  0.5809 &  0.5881      
             &          &  0.5388 &  0.5442     
             &          &  0.5762 &  0.5829   \\ 
 \ \ $T^2(R_n\ne R_p)$ & & 0.5810 &  0.5882      
             &          &  0.5389 &  0.5443      
             &          &  0.5763 &  0.5831   \\ 
\hline
 $S_V(T)$    &  24.940  &  25.506  &  27.294     
             &  29.655  &  30.210  &  31.956     
             &  27.382  &  27.980  &  29.873  \\ 
 \ \ $T$-indep.& 24.940 &  24.942  &  24.981     
             &  29.655  &  29.657  &  29.691    
             &  27.382  &  27.385  &  27.427  \\ 
 \ \ $T^2$ term &       &  0.5631 &  0.5783      
             &          &  0.5527 &  0.5662      
             &          &  0.5954 &  0.6115    \\ 
 \ \ $T^2(R_n\ne R_p)$ & & 0.5719 & 0.5871       
             &          &  0.5723 & 0.5859       
             &          &  0.5982 & 0.6145    \\ 
\hline
 $S_S(T)$    & --20.661  & --23.383 & --31.920     
             & --43.596  & --46.185 & --54.325     
             & --27.382  & --30.042 & --38.438  \\ 
 \ \ $T$-indep.&--20.661 & --20.674 & --20.832    
             & --43.596  & --43.606 & --43.760     
             & --27.382  & --27.393 & --27.571  \\ 
 \ \ $T^2$ term &       & --2.708  & --2.772       
             &          & --2.579  & --2.639       
             &          & --2.649  & --2.717    \\ 
 \ \ $T^2(R_n\ne R_p)$ & &--2.717  & --2.780       
             &          & --2.627  & --2.687      
             &          & --2.643  & --2.710    \\ 
\hline
\end{tabular}
\end{table}

It is of interest to compare the total symmetry energy using the coefficients given in 
Table \ref{encoef} for the two Skyrme interactions along the line of $\beta$ stability. 
In particular we calculate the symmetry energy $E_{sym} = S_V I^2 A + S_S I^2 A^{2/3}$ 
along a simplified stability line given by
\begin{eqnarray}
 \frac{Z_\beta}{A} &=& \frac{1}{2} \frac{1}{1 + \frac{1}{4} A^{2/3} \frac{a_c}{a_{sym}}}
\end{eqnarray}
which is determined by minimizing $E(Z,A) = E(A) + a_{sym} I^2 A + a_c Z^2/A^{1/3}$
with $a_c = 0.72$ MeV and $a_{sym} = 24$ MeV at $T = 0$. 
The $I_\beta = (N_\beta - Z_\beta)/A = (A - 2 Z_\beta)/A$ which is substituted 
into $E_{sym}(T) = S_V(T) I^2 A + S_S(T) I^2 A^{2/3}$ to give the 
symmetry energy along the line of $\beta$ stability. 
The energy is labeled 
 $E_{sym,\beta} = (S_V(T) A + S_S(T) A^{2/3}) I^2_\beta$
where $I_\beta = \left(\frac{A^{2/3}}{4} \frac{a_c}{a_{sym}}\right)/
   \left(1 + \frac{A^{2/3}}{4} \frac{a_c}{a_{sym}}\right)$ 
in Fig.\ref{fig1}. 
Fig.\ref{fig1} shows that the $T$ dependence for the symmetry energy is more 
visible for larger $A$ and the symmetry energy is larger for higher $T$. 
It shows also that the SKM($m^*=m$) parameter set has a symmetry energy
which is too small.

In Fig.\ref{fig2} the difference $\Delta E_{sym,\beta}$ between the symmetry energy
$E_{sym,\beta}(T)$ for SLy4 parameter set and the volume symmetry energy 
with $S_V = 24$ MeV and $S_S = 0$ (dotted curve in Fig.\ref{fig1}).
This figure shows that the symmetry energy is approximately $T$ independent
at around $A = 100$ with the total symmetry energy of 43 MeV for SLy4 set
compared to 46 MeV for pure volume symmetry energy only.
The symmetry energy for higher $T$ is smaller than the symmetry energy for lower $T$
at smaller $A$ than $A = 100$ while this result is opposite at larger $A$.
Differences of up to 4 MeV were found to be  present in the mid mass range of finite
nuclei. 
However for small nuclear masses and for a larger nuclear mass
($A \approx 200$ for $T =0$) the
two choices agree for the total symmetry energy for these coefficients.
This means SLy4 interaction has the total symmetry energy $E_{sym}(T)$
similar to the value for $E_{sym} = 24 I^2 A$ MeV 
up to $A \approx 200$ for $T < 3$ MeV.

\begin{figure}
                  
\includegraphics[width=5.0in]{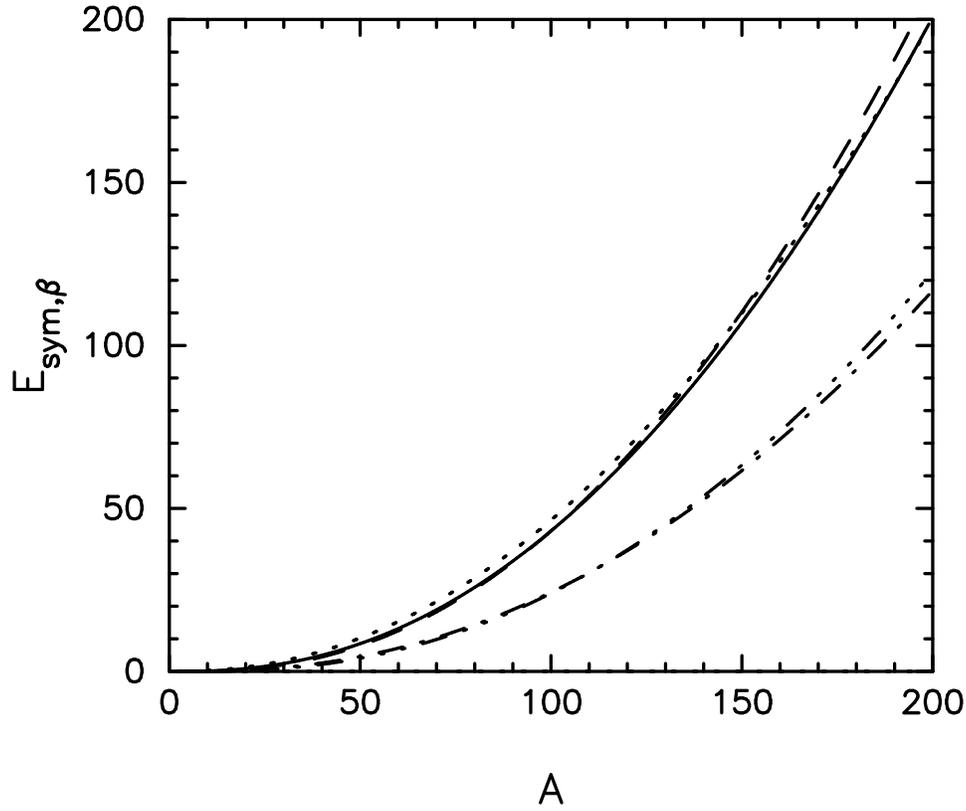}  

\caption{The total symmetry energy $E_{sym}(T) = S_V(T) I^2 A + S_S(T) I^2 A^{2/3}$
in MeV versus mass number $A$ along the line of stability. The 
upper curves are the interaction SLy4 (the solid curve for $T=0$ and the 
dashed curve for $T=3$ MeV) and the lower curves are SKM($m^*=m$) (the
dash-dotted curve for $T=0$ and the dash-dot-dot-dotted curve for $T=3$ MeV). 
The dotted curve is a pure volume term symmetry energy
with $S_V = 24$ MeV and $S_S =0$ MeV.
The $E_{sym,\beta}$ is in MeV.
  }    \label{fig1}
\end{figure}

\begin{figure}               
\includegraphics[width=5.0in]{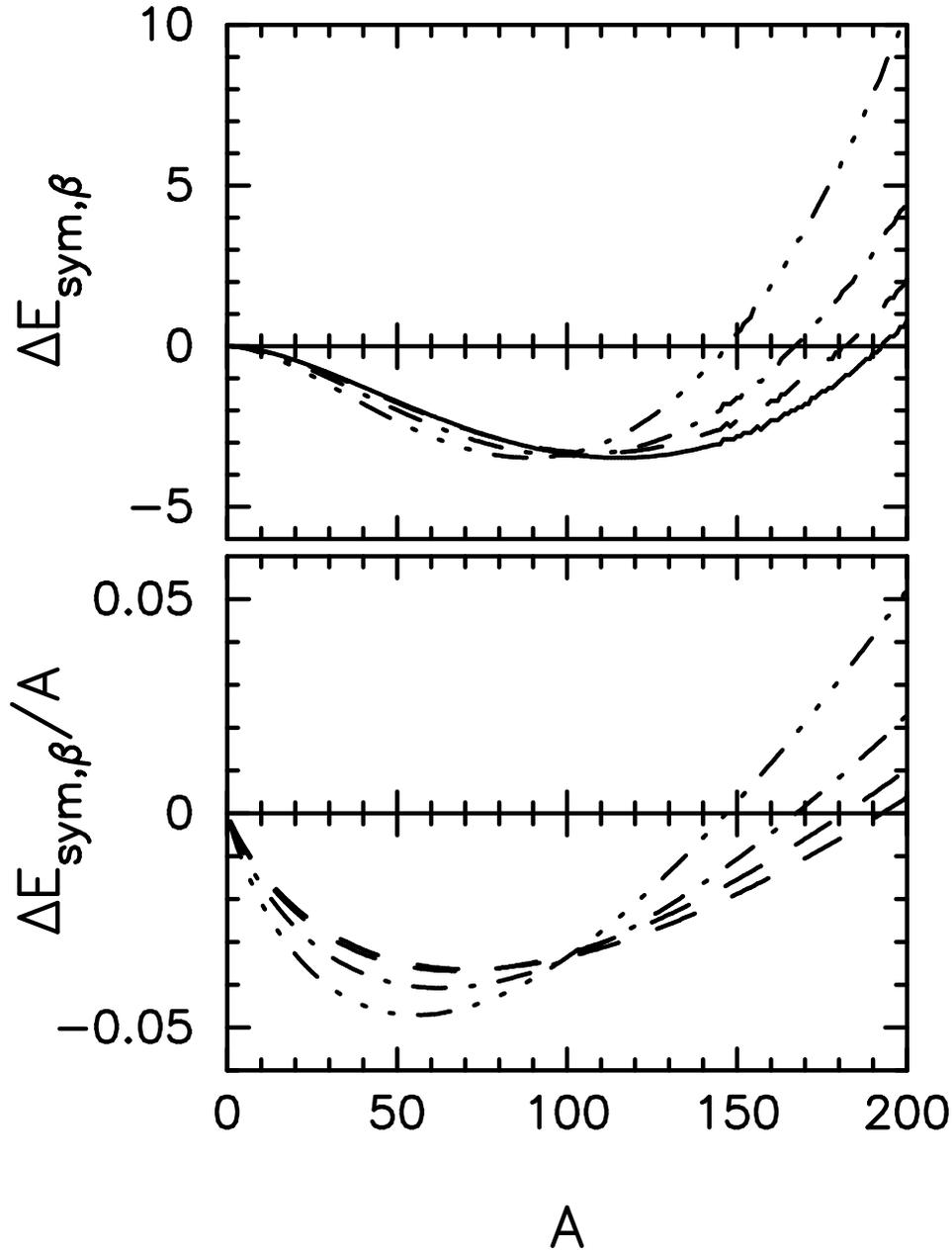}
 
\caption{Comparison of a symmetry energy with both volume and surface 
terms and a symmetry energy with just a volume term for SLy4 interaction. 
The symmetry energy is calculated along the line of $\beta$ stability. 
The upper and lower figures are the difference $\Delta E_{sym,\beta}$ 
and $\Delta E_{sym,\beta}/A$ in MeV between the two symmetry energies 
with both volume and surface symmetry energy terms for SLy4 interaction
and with volume symmetry energy term only (dotted curve in Fig\ref{fig1}).
The solid curve is at $T=0$, the dashed curve at $T=1$, 
the dash-dotted curve at $T=2$, and the dash-dot-dot-dotted curve at $T=3$ MeV.
At $A = 100$ and $T = 0$, the SLy4 interaction gives a total symmetry energy 
of 43.09 MeV, while the pure volume calculation has 46.44 MeV.  
   }   \label{fig2}
\end{figure}

\section{Conclusions}
 
 This paper presented an investigation of the energy of finite nuclei using density 
functional theory. A Skyrme approach for the nuclear interaction was used in this study. 
Bulk, surface, Coulomb and both volume and surface symmetry energies are calculated 
for a nucleus at both zero and non-zero temperatures. 
It is shown that the surface symmetry energy term is the most sensitive to 
the temperature while the bulk energy term is the least sensitive.
Understanding features associated 
with the volume and surface symmetry energy was the main part of this study. 
Specifically, the volume and surface symmetry energy coefficients $S_V$ and $S_S$ were 
calculated for two different Skyrme interactions, SLy4 and SKM($m^*=m$), and the 
results compared. The SLy4 interaction with a density dependent effective mass $m^*$ is 
in better agreement with known features of the symmetry energy of finite nuclei. 
The results suggest the importance of a density dependent effective mass. 
The results also show that the temperature $T$ dependence of the energy is
insensitive to the force used and even more insensitive
to the difference of $R_p$ and $R_n$.
These results are quite different than the $T = 0$ results which are
sensitive to the force used and the difference of $R_n$ and $R_p$
\cite{ref22,prl85}.

Then a comparison was made between the SLy4 Skyrme interaction results 
with $S_V(T)$ and $S_S(T)$ of Table \ref{encoef} 
($S_V = 31$ MeV, $S_S = -41$ MeV at $T = 0$)
and a model without surface effects $S_S = 0$ MeV
and with a volume coefficient $S_V = 24$ MeV. 
The results were compared in Figs.\ref{fig1} and \ref{fig2}. 
Differences of up to 4 MeV were found to be  present in the mid mass range of finite 
nuclei. However for small nuclear masses and for a larger nuclear mass 
($A \approx 200$ for $T =0$) the 
two choices agree for the total symmetry energy for these coefficients. 
The volume term $S_V = 31$ MeV in SLy4 is much larger than the typical 
textbook value of $S_V \approx 24$ MeV of the Weizsacker mass formula. 
The Weizsacker mass formulae with $S_V \approx 24$ MeV was fit to data 
without a surface term. Adding a surface term that reduces the symmetry 
energy will result in an increase in the volume coefficient.
Various constraints give limits \cite{ref20,ref21,ref22} on $S_V$ with in 
the range $24 {\rm MeV} \le S_V \le 33 {\rm MeV}$. 
Understanding the volume coefficient in finite nuclei is necessary in 
extrapolations of the symmetry energy to neutron star physics.  

This work supported by Basic Science Research Program through
NRF of Korea funded by the Ministry of
Education, Sciences and Technology under Grant 2009-0072500
and by Department of Energy under Grant DE-FG02ER-40987.

\appendix

\section{} 

For a density $\rho(\vec r) = \rho_c/[1 + e^{(r-R)/a}]$,
using partial integral,
\begin{eqnarray}
 \int d^3 r \rho(\vec r)^n 
  &=& 4\pi \rho_c^n \int_0^\infty \frac{r^2 dr}{[1 + e^{(r-R)/a}]^n}
   = \frac{4\pi}{3} \rho_c^n \int_0^\infty \frac{r^3 dr}{a} 
       \frac{n e^{(r-R)/a}}{[1 + e^{(r-R)/a}]^{n+1}}        \nonumber \\
  &=& \frac{4\pi}{3} \rho_c^n \int_{-\infty}^\infty dy (ay+R)^3
       \frac{n e^y}{(1 + e^y)^{n+1}}
\end{eqnarray}
For the last expression, the lower limit of the integration of $y$ is extended
from $-R/a$ to $-\infty$ which is a good approximation for $n \ge 1$ 
since the symmetric factor $e^y/(1 + e^y)^2$ peaks at $y=0$ and becomes 
zero at $\pm\infty$.
The integral part for various $n$ are
\begin{eqnarray}
 \int_{-\infty}^\infty dy (ay+R)^3 \frac{e^y}{(1 + e^y)^2}
  &=& \left(\pi^2 a^2 R + R^3\right)     \\
 \int_{-\infty}^\infty dy (ay+R)^3 \frac{2 e^y}{(1 + e^y)^3}
  &=& \left(-\pi^2 a^3 + \pi^2 a^2 R - 3 a R^2 + R^3\right)  \\
 \int_{-\infty}^\infty dy (ay+R)^3 \frac{3 e^y}{(1 + e^y)^4}
  &=& \left(-\frac{3\pi^2}{2} a^3 + (3+\pi^2) a^2 R - \frac{9}{2} a R^2 
            + R^3\right) \\
 \int_{-\infty}^\infty dy (ay+R)^3 \frac{4 e^y}{(1 + e^y)^5}
  &=& \left(-\frac{(6+11\pi^2)}{6} a^3 + (6+\pi^2) a^2 R 
            - \frac{11}{2} a R^2 + R^3\right)     \\
 \int_{-\infty}^\infty dy (ay+R)^3 \frac{5 e^y}{(1 + e^y)^6}
  &=& \left(-\frac{5(6+5\pi^2)}{12} a^3 + \frac{(35+4\pi^2)}{4} a^2 R 
           - \frac{25}{4} a R^2 + R^3\right)        \\
 \int_{-\infty}^\infty dy (ay+R)^3 \frac{(1/3) e^y}{(1 + e^y)^{4/3}}
  &=& \left(159.3784 a^3 + 54.8029 a^2 R + 7.66445 a R^2 + R^3\right)   \\
 \int_{-\infty}^\infty dy (ay+R)^3 \frac{(2/3) e^y}{(1 + e^y)^{5/3}}
  &=& \left(15.855575 a^3 + 15.77375 a^2 R + 2.2230575 a R^2 + R^3\right)   \\
 \int_{-\infty}^\infty dy (ay+R)^3 \frac{(4/3) e^y}{(1 + e^y)^{7/3}}
  &=& \left(-5.0303125 a^3 + 8.81615625 a^2 R -1.335546875 a R^2 + R^3\right) \\
 \int_{-\infty}^\infty dy (ay+R)^3 \frac{(5/3) e^y}{(1 + e^y)^{8/3}}
  &=& \left(-7.80506 a^3 + 9.10458 a^2 R - 2.276943 a R^2 + R^3\right)   \\
 \int_{-\infty}^\infty dy (ay+R)^3 \frac{(7/3) e^y}{(1 + e^y)^{10/3}}
  &=& \left(-11.6424 a^3 + 10.81947 a^2 R - 3.58554 a R^2 + R^3\right)   \\
 \int_{-\infty}^\infty dy (ay+R)^3 \frac{(8/3) e^y}{(1 + e^y)^{11/3}}
  &=& \left(-13.26781 a^3 + 11.836907 a^2 R - 4.07693333 a R^2 + R^3\right)   \\
 \int_{-\infty}^\infty dy (ay+R)^3 \frac{(7/6) e^y}{(1 + e^y)^{13/6}}
  &=& \left(-3.00561 a^3 + 9.07054 a^2 R - 0.7352638484 a R^2 + R^3\right)   \\
 \int_{-\infty}^\infty dy (ay+R)^3 \frac{(13/6) e^y}{(1 + e^y)^{19/6}}
  &=& \left(-10.7804 a^3 + 10.331 a^2 R - 3.30669 a R^2 + R^3\right)
\end{eqnarray}
For $n=1$, we get the constraint
\begin{eqnarray}
 A &=& \int d^3 r \rho(\vec r)
    = \frac{4\pi}{3} \rho_c \int_{-\infty}^\infty dy
        \frac{(ay+R)^3 e^y}{(1 + e^y)^2}
    = \frac{4\pi}{3} R^3 \rho_c 
        \left[1 + \pi^2 \left(\frac{a}{R}\right)^2\right]
\end{eqnarray}
This relation expresses the central density $\rho_{pc}$ and $\rho_{nc}$
in terms of the size $R$ and a given diffuseness parameters $a$ 
for a given number of protons $Z$ and neutrons $N$.
Similarly for the gradient dependent term,
\begin{eqnarray}
 \int d^3 r \rho(\vec r) \nabla^2 \rho(\vec r)
  &=& 4\pi \int_0^\infty r^2 dr \left[\frac{\rho_c}{1 + e^{(r-R)/a}}\right]
      \frac{1}{r} \frac{d^2}{dr^2} r \left[\frac{\rho_c}{1 + e^{(r-R)/a}}\right]
           \nonumber  \\
  &=& 4\pi \rho_c^2 \int_{-\infty}^\infty d y \frac{1}{(1 + e^y)^2}
    \left[2 \frac{(a y + R)^2}{a} \left(\frac{e^y}{1 + e^y}\right)^2
       - \frac{(a y + R)^2}{a} \left(\frac{e^y}{1 + e^y}\right) 
              \right.   \nonumber  \\   & & \hspace{1.5in}  \left.
       - 2 {(a y + R)} \left(\frac{e^y}{1 + e^y}\right)\right]
           \nonumber  \\
  &=& - \frac{4 \pi}{3} R^3 \frac{\rho_c^2}{2Ra}
              \left[1 + \left(\frac{\pi^2}{3} - 2\right) 
              \left(\frac{a}{R}\right)^2 \right]
\end{eqnarray}


\begin{thebibliography}{99}

\bibitem{ref1}  C.F.Weizsacker, Z.phys. 96, 431 (1935)
\bibitem{ref2}  H.A.Bethe and R.F.Bacher, Rev. Mod. Phys. 8, 82 (1936)
\bibitem{ref3}  W.D.Meyers and W.J.Swiatecki, Nucl. Phys. A 81, 1 (1966)
\bibitem{ref4}  W.D.Meyers and W.J.Swiatecki, Ann. Phys. 55, 395 (1969)
\bibitem{ref5}  W.D.Meyers and W.J.Swiatecki, Ann. Phys. 84, 186 (1974)
\bibitem{ref6}  W.D.Meyers, Droplet Model of Atomic Nuclei, Plenum, New York, 1975 
\bibitem{ref7}  A.Bohr and B.R.Mottelson, Nuclear Structure, Vol I, Benjamin, New York 1969
\bibitem{ref8}  G.F.Bertsch and A.Z.Mekjian, Ann. Rev. Nucl. Sci. V22, 25 (1972)
\bibitem{ref9}  A.M.Lane and A.Z.Mekjian, Advances in Nuclear Physics, V7, 97 (1973) 
       (Eds. M.Baranger and E.Vogt)
\bibitem{ref10} S.Das Gupta, A.Z.Mekjian and B.Tsang, Advances in Nuclear Physics, 
       V26, 89 (2001) (Eds. J.Negele and E.Vogt,)
\bibitem{ref11} C.B.Das, S.Das Gupta, W.B.Lynch, A.Z.Mekjian and B.Tsang, Phys. 
       Reports 406, 1 (2005)  
\bibitem{ref12} C.B.Das, S.Das Gupta and A.Z.Mekjian Phys. Rev. C67, 064607 (2003) 
\bibitem{ref13} H.S. Xu,etal, Phys. Rev. Lett 85, 716 (2000) 
\bibitem{ref14} B.-A. Li, Phys. Rev. Lett. 85, 4221 (2000)
\bibitem{ref15} S.J. Lee and A.Z. Mekjian, Phys. Rev. C77, 054612 (2008)
\bibitem{ref16} S.J. Lee and A.Z. Mekjian, Phys.Lett. 580, 137 (2004) 
\bibitem{ref17} S.J. Lee and A.Z. Mekjian Phys. Rev. C68, 014608 (2003) 
\bibitem{ref18} B.-A. Li and U.Schroder (Eds), Isospin Physics in Heavy Ion Collisions 
        at Intermediate Energies, Nova Science, New York, 2001 
\bibitem{ref19} P. Danielewicz, R.Lacey and W.B.Lynch, Science 298, 1592 (2003)
\bibitem{ref20} P. Danielewicz, Nucl. Phys. A 727, 233 (2003)  
\bibitem{ref21} P. Danielewicz and J.Lee, Int. J. Mod. Phys. E18, 892 (2009) 
\bibitem{ref22} A.W.Steiner, M.Prakash, J.M.Lattimer and P.J.Ellis, Phys. Reports 411, 325 (2005) 
\bibitem{prc79} S.J. Lee and A.Z. Mekjian, Phys. Rev. C79, 044323 (2009)

\bibitem{dan9} H.A. Bethe, R.F. Bacher, Rev. Mod. Phys. 8, 82 (1936).
\bibitem{dan11} M.A. Preston, R.K. Bhaduri, {\bf Structure of the Nucleus}, 
         Addison.Wesley, Reading, MA, 1975.
\bibitem{dan10} W.D. Myers, W.J. Swiatecki, Nucl. Phys. A 81, 1 (1966).
\bibitem{dan14} W.D. Myers, W.J. Swiatecki, Ann. Phys. 55, 395 (1969).
\bibitem{dan15} W.D. Myers, W.J. Swiatecki, Ann. Phys. 84, 186 (1974).
\bibitem{dan16} W.D. Myers, {\bf Droplet Model of Atomic Nuclei}, IFI/Plenum, 
        New York, 1975.

\bibitem{skmst}E. Chabanat, P. Bonche, P. Haensel, J. Meyer, and R. Schaeffer,
        Nucl. Phys. {\bf A635}, 231 (1998).

\bibitem{npa818} P. Danielewicz and J. Lee, Nucl. Phys. {\bf A818}, 36 (2009).
\bibitem{prl85} B.A. Brown, Phys. Rev. Letts. 85, 5296 (2000).

\bibitem{skt8}F. Tondeur, M. Brack, M. Farine, and J.M. Pearson,
        Nucl. Phys. {\bf A420}, 297 (1984).
\bibitem{ski3}P.-G. Reinhard and H. Flocard, Nucl. Phys. {\bf A584}, 467 (1995).

\end{thebibliography}
\end{document}